\documentstyle [12pt,a4wide]{article}
\def\mom{{\cal H}_i }

\def\xxp{(x\leftrightarrow x')}
\def\xxpp{(ix\leftrightarrow jx')}

\begin{document}  

\thispagestyle{empty}
\setcounter{page}{1}
\begin{titlepage}

\begin{flushright}
Imperial/TP/97--98/17.
\end{flushright}
\vskip 1cm
\begin{center}
{\large{\bf  A classical history theory: Geometrodynamics and
general field dynamics regained. }}   
\vskip 2cm
{ I.~Kouletsis
\vskip 0.4cm {\it Theoretical Physics Department\\
Blackett Laboratory\\
Imperial College of Science, Technology \& Medicine\\
London  SW7 2BZ, U.K.\\E-mail address: tpmsc17@ic.ac.uk. }}
\vskip 0.7cm
\end{center}

\vskip 0.9cm
\begin{abstract}
Assuming that the Hamiltonian of a canonical field theory can be
written in the form  $N H + {N^i} {H_i}$, and using as the only input the  
actual choice of the canonical variables, we derive: (i) The algebra
satisfied by $H$ and $H_i$, (ii) any constraints, and (iii) the most
general canonical representation for $H$ and $H_i$. This completes
previous work by Hojman, Kucha\v{r} and Teitelboim who had to impose a set of
additional postulates, among which were the form of the canonical
algebra and the requirement of path-independence of the dynamical evolution. 
A prominent feature of the present approach is the replacement of the
equal-time Poisson bracket with one evaluated at general times. The
resulting formalism is therefore an example of a classical history theory---an
interesting fact, especially in view of recent work by Isham {\it et al}.      

\end{abstract}
\end{titlepage}
\renewcommand{\theequation}{\thesection.\arabic{equation}}
\let\ssection=\section
\renewcommand{\section}{\setcounter{equation}{0}\ssection}

\section{Introduction.}
We discuss the issue concerning the transition from a general
canonical Hamiltonian of the form $NH + {N^i} {H_i}$ to a specific
canonical representation. The form of the Hamiltonian is general
enough to incorporate a large variety of canonical field
theories---including general relativity---while the information that
distinguishes one theory from the other comes, solely, through the choice
of canonical variables. Changing the geometric interpretation of the
functions $N$ and $N^i$ changes the form of the algebra of the
canonical generators $H$ and $H_i$.

Some of this has been discussed already by Hojman, Kucha\v{r} 
and Teitelboim\cite{HKT} who succeeded in deriving the canonical form of 
covariant spacetime theories from a few simple postulates. 
As stated by the authors themselves, however, a reduction of these
postulates to the minimum was not attempted and some  
redundancy was left in the system. A couple of superfluous
requirements were pointed out at the end of their paper and
were not used in a subsequent one\cite{K} but, still, the exact
relationship between the remaining postulates was not clarified
completely, and a further reduction seemed to be possible.
We show that this is indeed the case, and that one can derive the 
complete set of postulates in \cite{HKT} from just the minimum requirement
that the canonical Hamiltonian is of the form $NH + {N^i} {H_i}$.

Just as interesting as this result, however, is the ensuing 
conclusion that the reduction of the postulates to the minimum could
never have been achieved in the framework used by Kucha\v{r} {\it et
al} due to their use of equal-time Poisson brackets. This is because, in
an equal-time formalism, Poisson brackets that involve the time derivatives of
the canonical variables cannot be defined---at least, not without the
addition of further structure. Seen from a spacetime perspective,
however, these brackets ought to be treated in an equivalent way, in
which case they would give important information about the theory's
kinematics. In the equal-time formalism the missing information is
precisely recovered by the additional postulates imposed in
\cite{HKT}, most notably that of the Dirac algebra.

The present approach, on the other hand, is based on a Hamiltonian 
formalism whose phase space includes the fields at general times,
i.e., is defined over the space of {\it classical histories}. An exact
correspondence with the spacetime picture is thus established
from the beginning and the reduction of the
postulates comes as a direct consequence.
In fact, the effectiveness of the history formalism could suggest that the
latter is genuinely superior to its equal-time counterpart, although   
this depends on whether it is possible, or not, to establish a
direct link between the equal-time and the history  
approaches. This is also discussed by Isham {\it
et al}\cite{Isham} in the context of continuous-time histories.

This paper is organized as follows. In section 2 we review the 
existing work on the subject and emphasize the main issues. 
The aim of this section is twofold: first, to act as an introduction
for the reader who is not familiar with the subject and, second,
to give the motivation for the construction that follows.  
In section 3 we present the framework for passing from the 
Lagrangean to a Hamiltonian formalism defined over the space of
classical histories. The formalism incorporates both  
constrained and unconstrained systems and---at least for the issues of
interest---is a simpler alternative to the Dirac method.

In section 4 we apply the history formalism to transform the postulated form 
$N H + N^i H_i$ of the canonical Hamiltonian to a set of kinematical
conditions on the canonical generators and, in section 5, we use these
conditions to derive the additional postulates imposed by Kucha\v{r}
{\it et al}. Having established the connection between our approach and the
approach in \cite{HKT} we can be certain that the expression $N H + N^i
H_i$, alone, is enough to determine the system. This includes the
canonical algebra, any constraints, as well as the most
general canonical representation of the generators.

\section{Motivation.}
The canonical decomposition of Hilbert's action brings the theory of
general relativity into the Hamiltonian form\cite{ADM}
\begin{equation}
{\cal S} = \int d^3x dt [ p^{ij} {\dot g}_{ij} - N H^{gr} - N^i H^{gr}_i ].
\label{eq:1}
\end{equation}
The lapse function $N$ and the shift vector $N^i$ acquire the meaning
of Lagrange multipliers and, as a result, the canonical generators 
\begin{eqnarray}
&& H^{gr} = {1 \over 2} (g_{ik} g_{jl} + g_{il} g_{jk} - g_{ij}
g_{kl}) p^{ij} p^{kl} - g^{1 \over 2} {^{(3)}}R + g^{1 \over 2}
\Lambda ,
\label{eq:Hm}
\\
&& H^{gr}_i = -2 {{p_i}^j}_{|j},
\label{eq:Mm}
\end{eqnarray}
are constrained to vanish:
\begin{eqnarray}
&& H^{gr} \simeq 0 ,
\label{eq:Hconst}\\
&& H^{gr}_i \simeq 0 .
\label{eq:Mconst}
\end{eqnarray}
The generators can be shown to satisfy the Dirac algebra\cite{Dirac},
\begin{eqnarray}
&& \{H(x), H (x')\} = g^{ij}(x){H_i}(x)
\delta_{,j}(x,x')-\xxp\label{eq:D1} 
\\
&& \{H(x), {H_i}(x')\} = H(x) \delta_{,i}(x,x') + {H}_{,i}(x) \delta (x,x')
\label{eq:D2}
\\
&& \{{H_i}(x),{H_j}(x')\} = {H_j}(x)\delta_{,i}(x,x')-\xxpp ,
\label{eq:D3}
\end {eqnarray}
that is also satisfied by the canonical generators of a parametrized field 
theory. For a field theory that is not parametized, the
canonical algebra is not very different from the Dirac one 
and can always be recognized as a suitably modified version of it.

\paragraph{The Dirac algebra and the principle of path independence.}
This universality implies that the Dirac algebra is connected
with a very general geometric property of spacetime which is
independent of the specific dynamics of the canonical theory. 
The fact that the Dirac algebra is merely a kinematical consistency 
condition was shown by Teitelboim\cite{T}, who proceeded to derive it from a 
simple geometric argument corresponding to the integrability of
Hamilton's equations.
 
This consistency argument, termed by Kucha\v{r}\cite{KBT} ``the
principle of path independence of the dynamical evolution'', 
ensures that the change in the canonical variables during the evolution 
from a given initial surface to a given final surface is independent
of the particular sequence of intermediate surfaces used in the actual
evaluation of this change. 

To be precise, besides the assumption of path independence---which  
applies regardless of the specific form of the canonical
Hamiltonian---Teitelboim's derivation also explicitly involved the
assumption that the Hamiltonian is decomposable according to the lapse-shift
formula written in equation (\ref{eq:1}). Using these two
postulates, together, he was then led to the conclusion that in order
for the theory to be consistent 
the phase space should be restricted by the initial value equations 
(\ref{eq:Hconst}-\ref{eq:Mconst}) while the canonical generators
should satisfy the Dirac algebra (\ref{eq:D1}-\ref{eq:D3}).  

Strictly speaking, the very last statement is not true 
due to a mistake in Teitelboim's reasoning concerning the fact that the
system is constrained. 
The correct algebra---as it arises from the requirement of path
independence---is nonetheless very similar 
to the Dirac one but modified by certain terms $G$, $G_i$ and
$G_{ij}$ whose first partial derivatives with respect to the
canonical variables vanish on the constraint surface: 
\begin{eqnarray}
&& \{ H (x), H (x') \} = g^{ij}(x){H}_i(x)
\delta_{,j}(x,x') + G(x,x') -\xxp ,
\label{eq:E1} \\
&& \{ H(x), {H}_i(x') \} = H(x) \delta_{,i}(x,x') + H_{,i}(x) \delta
(x,x') + G_i(x,x') ,
\label{eq:E2} \\
&& \{{H}_i(x),{H}_j (x')\} = {H}_j(x)\delta_{,i}(x,x') + G_{ij}(x,x') -\xxpp . 
\label{eq:E3}
\end {eqnarray}
A derivation of the above set of relations---which we call the weak
Dirac algebra---is given in section 5.

\paragraph{The problem of deriving a physical theory from just the
canonical algebra.} 
The principle of path independence was an indication 
that a Hamiltonian theory need not be based exclusively on the canonical 
decomposition of some given spacetime action, but may also have an 
independent status. However, one finds that some of the
necessary information is missing when one tries to construct specific
canonical theories via the principle of path independence
alone. The reason is that the weak Dirac algebra---which  
expresses the principle in the canonical language---allows a vast
variety of representations to exist whose physical relevance is doubtful. 

To give an example, we consider the case when the canonical 
variables are the spatial metric and its conjugate momentum,
and we take the limit of the algebra (\ref{eq:E1}-\ref{eq:E3})
when all the terms $G$, $G_i$ and $G_{ij}$ are identically zero, i.e.,
we take the usual Dirac case.
We let the $H_i$ generator be the super-momentum of the 
gravitational field,
\begin{equation}
\mom(x) = H^{gr}_i(x), 
\label{eq:mome}
\end{equation}
and require that the normal generator $H(x)$ be a scalar density of weight one.
Under these conditions, the second and third Dirac relations
(\ref{eq:D2}-\ref{eq:D3}) are 
satisfied, and the Dirac algebra---which can be seen as a set of 
coupled differential equations for the canonical
generators---decouples completely. One is left with a single first-order 
equation for $H(x)$, equation (\ref{eq:D1}), which is
normally expected to admit an infinite number of distinct solutions.

In particular, one can take $H(x)$ to have the form
\begin{equation}
H(x) = g^{1 \over 2} W[h,f](x),
\label{eq:myD}
\end{equation}
where the weight-zero quantities $h$ and $f$ are defined\cite{M} by
\begin{eqnarray}
&& h = g^{-{1 \over 2}} H^{gr} ,
\nonumber\\
&& f = g^{-1} g^{ij} H^{gr}_i H^{gr}_j .
\label{eq:hf}
\end{eqnarray}
The resulting equation for the function $W[h,f]$ is
\begin{equation}
{1 \over 2}W{{\partial W} \over {\partial f}} = f({{\partial W} \over
{\partial f}})^2 - {1 \over 4}({{\partial W} \over {\partial h}})^2 +
{1 \over 4} ,
\label{eq:eksisosh}
\end{equation}
and can be shown to admit a family of solutions that is
parametrized by an arbitrary function of one variable\cite{IKDID}.

The super-Hamiltonian of general
relativity, arising when $W[h,f]=h$, is the only one of these
solutions that is ultralocal in the field momenta. The ultralocality is
actually related to the geometric meaning of the canonical variables,
but this will be discussed properly in section 5. For the moment,
note that if one uses the weak Dirac algebra
(\ref{eq:E1}-\ref{eq:E3}) as the starting point of the above
calculation---which is the correct thing to do---one is forced
to solve a set of coupled differential equations whose actual form is unknown!

\paragraph{Selecting the physical representations of the Dirac algebra.}
Deriving geometrodynamics from plausible first principles,
Hojman, Kucha\v{r} and Teitelboim\cite{HKT} chose to lay the stress on the
concept of infinite dimensional groups, and placed the strong Dirac algebra
at the centre of their approach. They expected that the closing
relations (\ref{eq:D1}-\ref{eq:D3}) themselves carry enough 
information about the system to uniquely select a physical
representation, but they were unable to extract this
information directly from them. We now know that the existence of
solutions like (\ref{eq:myD}) was the reason why. 

What the authors of \cite{HKT} did instead, was to follow an indirect route and
select the physically relevant representations by  
supplementing the strong Dirac algebra with four additional conditions. 
Specifically, they introduced the tangential and normal generators of
hypersurface deformations, defined respectively by
\begin{eqnarray}
&& {H^D}_i(x) := {{\cal X}^{\alpha}}_i(x) {\delta \over {\delta 
{\cal X}^{\alpha}}}(x),
\label{eq:defo1}
\\
&& {H^D}(x) := {n^{\alpha}}(x) {\delta \over {\delta {\cal X}^{\alpha}}}(x),
\label{eq:defo2}
\end{eqnarray}
and acted with these on the spatial metric:
\begin{eqnarray}
&& {H^D}_k(x')g_{ij}(x) = g_{ki}(x) {\delta}_{,j}(x,x') + 
g_{kj}(x) {\delta}_{,i}(x,x') + g_{ij,k}(x) {\delta}(x,x'),
\label{eq:extra1}
\\
&& H^D(x')g_{ij}(x) = 2 n_{{\alpha};{\beta}}(x) {{\cal X}^{\alpha}}_i(x)
{{\cal X}^{\beta}}_j(x) \delta(x,x').
\label{eq:extra2}
\end{eqnarray}

Then, they required that equations
(\ref{eq:extra1}-\ref{eq:extra2})---which are purely kinematical and
hold in an arbitrary Riemannian spacetime---should also be satisfied
by the canonical generators, 
\begin{eqnarray}
&& \{ g_{ij}(x),H_k(x') \} = g_{ki}(x) {\delta}_{,j}(x,x') + 
g_{kj}(x) {\delta}_{,i}(x,x') + g_{ij,k}(x) {\delta}(x,x'), 
\label{eq:can1}
\\
&& \{ g_{ij}(x),H(x') \} \propto {\delta}(x,x'),
\label{eq:can2}
\end{eqnarray}
so that any dynamics in spacetime would arise as a different canonical
representation of the universal kinematics. Note that only the
ultralocality of the second Poisson bracket was actually used. 
The justification and geometric interpretation of equations
(\ref{eq:can1}) and (\ref{eq:can2}) can be found in \cite{HKT}.

The strong Dirac algebra with the conditions
(\ref{eq:can1}), (\ref{eq:can2}) 
results in a unique representation for the generators $H$ and
$H_i$, corresponding to the super-Hamiltonian (\ref{eq:Hm}) and
super-momentum (\ref{eq:Mm}) of general relativity. 
The requirement of path independence---which was imposed as an
additional postulate to the algebra---enforces the initial value constraints
(\ref{eq:Hconst}-\ref{eq:Mconst}) and, hence, the complete set of  Einstein's 
equations is recovered. The most general scalar field Lagrangean with
a non-derivative coupling to the metric was derived along similar lines\cite{K}.

The precise assumptions used by the authors were
summarized at the end of their paper. They are written here 
in an equivalent form and, in the case of pure gravity, they are the
following:    

(i) The evolution postulate: The dynamical evolution of the theory is
generated by a Hamiltonian that is decomposed according to
the lapse-shift formula, equation (\ref{eq:1}).

(ii) The representation postulate: The canonical generators must 
satisfy the closing relations (\ref{eq:D1}-\ref{eq:D3}) of the strong
Dirac algebra. 

(iii) Initial data reshuffling: The Poisson bracket (\ref{eq:can1})
between the super-Hamiltonian and the configuration variable $g_{ij}$
must coincide with the kinematical relation (\ref{eq:extra1}). 

(iv) Ultralocality: The Poisson bracket (\ref{eq:can2}) between the 
super-momentum and the configuration variable $g_{ij}$ must
coincide with the kinematical relation (\ref{eq:extra2}).

(v) Reversibility: The time-reversed spacetime must 
be generated by the same super-Hamiltonian and super-momentum as the 
original spacetime.

(vi) Path independence: The dynamical evolution predicted by the
theory must be such that the change in the canonical variables during
the evolution from a given initial surface to a given final one is
independent of the actual sequence of intermediate surfaces used in
the evaluation of this change.

\paragraph{The need for a detailed understanding of the selection postulates.}
The above assumptions comprise a set of natural first principles on 
which the canonical formulation of a theory can be based. There is a certain
sense, however, in which they are not completely satisfying. First, they do not
correspond to a minimum set and, second, the connection between them
is not very clear. The authors themselves pointed out the 
redundancy of the reversibility postulate (v) as well as the fact that the
third closing relation of the representation postulate (ii) is made
redundant by the reshuffling requirement (iii). They stressed the
need for understanding the precise reason why some equations hold strongly
while others hold only weakly and, in particular, for clarifying the
relationship between the strong representation postulate (ii) and the
weak requirement of path independence (vi).   

The revised form of Teitelboim's argument makes such a
clarification an even more important issue since, now, the strong
representation requirement---which is at the very heart of the
approach in \cite{HKT}---seems to be unjustified. Adding to that, one
can repeat 
Teitelboim's argument in the reverse order and show that the dynamical
evolution of the theory should also hold weakly, in contrast to the
strong equalities in postulates (iii) and (iv). On the other hand, we already
know that any attempt to replace these equalities by weak ones 
would result in a situation where the actual form of the
differential equations would not be known and no further progress would
be made. Even if one justifies postulates (ii), (iii) and (iv) by
assuming that general relativity just happens to exist on the strong
limit of path independence, one will not be able to justify postulate
(vi) whose weak imposition is necessary in order for the theory to be
consistent.

Putting the issue of the weak equalities aside, an understanding of the 
exact relationship between the postulates is also needed if the method of
\cite{HKT} is to be applied to the case of an arbitrary canonical
algebra. The reason is that, in the existing formulation of the postulates,
the overall consistency is only made certain by the fact that the
reshuffling and ultralocality assumptions (iii) and (iv) are respected by
the dynamical law of the theory (i). On the other hand, nothing in the
remaining postulates ensures that assumptions (iii) and (iv) are the
only ones compatible with this law. If different compatible
assumptions are used as supplementary conditions to the algebra, the
above method will yield different canonical representations. Note,
however, that the dynamical law of the theory is the only
assumption---besides the principle of path independence---that enters
the derivation and geometric interpretation of the algebra. 
It follows that if the existing formulation of the postulates is used
as an algorithm for passing from the interpretation of an algebra
to its physical representations, it will be highly ambiguous.

We basically have in mind the interpretation of the 
genuine Lie algebra that was discovered by Brown and
Kucha\v{r}\cite{BK}. There have been some interesting approaches in this
subject\cite{M}\cite{KR}\cite{IK1}, but they all have revolved round the
abstract algebra, thus ignoring the actual procedure that led from
the Dirac algebra to general relativity.  For example, the solutions
found by Markopoulou\cite{M} are essentially the equivalent of the
solutions (\ref{eq:myD}) of the Dirac algebra. They do not depend on
anything but the algebra and, as such, they are expected to
contain certain unphysical representations among them. An unambiguous
formulation of the algorithm in \cite{HKT} will find here
a most natural application.

As a final note, we point out an asymmetry in the formulation of the
postulates that actually provides the main
motivation for the paper. It concerns the kinematic equations
(\ref{eq:extra1}) and (\ref{eq:extra2}) on
which postulates (iii) and (iv) are based. Namely, if the identification
of the canonical generators with the generators of normal and
hypersurface deformations is to be taken as a fundamental principle
in the canonical theory, one anticipates that it will hold for both the
canonical variables. However, in an equal-time formalism one can neither
confirm nor reject this conjecture simply because the action of the
deformation generators on the canonical  momenta cannot be defined
without additional structure. Marolf\cite{Marolf} used the Hamiltonian as an
additional structure to extend the Poisson bracket from a Lie bracket on phase
space to a Lie bracket on the space of histories. What we do,
instead, is to ignore completely the equal-time formalism, and proceed
with a phase space whose Poisson bracket is defined over the space of
histories from the beginning.

\section{The history formalism.}

\subsection{The unconstrained Hamiltonian.}
Consider the theory described by the canonical action
\begin{eqnarray} 
&& S[q^A , p_A] = \int d^3dt \bigg( p_A \dot{q^A} - {\cal H} \bigg),
\nonumber\\
&& {\cal H} = N H + N^i H_i,
\label{eq:UnA}
\end{eqnarray}
where $N$ and $N^i$ are prescribed functions of space and time.
The generators $H$ and $H_i$ are given functions of the canonical fields
$(q^A,p_A)$ and may also depend on additional prescribed fields $c^K$. 
The index $A$ runs from 1 to half the total number of canonical
variables, while $K$ runs from 1 to the total number of prescribed
fields.

One can generalize the phase space to include the canonical fields at
all times by introducing the space of histories,
\begin{equation}
\bigg( q^A(x,t) , p_A(x,t) \bigg),
\label{eq:histoir}
\end{equation}
and defining on it the Poisson bracket 
\begin{equation}
\{ q^A(x,t) , p_B(x',t') \} = {{\delta}^A}_B {\delta}(x,x') {\delta}(t,t').
\label{eq:UnBr}
\end{equation}
The quantum analogue of the canonical fields in (\ref{eq:histoir}) is the
one-parameter family of Schr\"{o}edinger operators introduced by Isham
{\it et al} in their study of continuous time consistent
histories\cite{Isham}\cite{CJI}.

Using the bracket (\ref{eq:UnBr})---which turns the space of
histories into a Poisson manifold---the variation of the canonical
action can be concisely written in the form
\begin{eqnarray}
&& \{ S , q^A(x,t) \} \simeq 0
\label{eq:UnV1}
\\
&& \{ S , p_A(x,t) \} \simeq 0,
\label{eq:UnV2}
\end{eqnarray}
and defines a constraint surface on this space. The physical
fields are defined to satisfy these relations for each value of $x$ and $t$. 
For the particular form (\ref{eq:UnA}) of the canonical action, the
weak equations (\ref{eq:UnV1}-\ref{eq:UnV2})
become\footnote{Throughout this paper, the
functional derivative ${{\delta {\cal F}}\over{\delta q^A }}$ is
defined by
${{\delta {\cal F}}\over{\delta q^A }} =  
{{\partial {\cal F}}\over{\partial q^A }} + 
{{\partial {\cal F}}\over{\partial {q^A}_{,i} }}
{\partial}_{i} + {{\partial {\cal F}}\over{\partial {q^A}_{,ij} }}
{\partial}_{ij} + ...etc$. We will call ${\cal F}$ a functional,
although we essentially mean a local function of the canonical variables and
a finite number of their derivatives.}   
\begin{eqnarray}
&& \dot{q^A}(x,t) \simeq \int d^3x'dt' \{ q^A(x,t) , {\cal H}(x',t') \}
\equiv \int d^3x' {{\delta {\cal H}}\over{\delta p_A }}(x',t) \delta(x,x')
\label{eq:UnH1}
\\
&& \dot{p_A}(x,t) \simeq \int d^3x'dt' \{ p_A(x,t) , {\cal H}(x',t') \}
\equiv \int d^3x' {{\delta {\cal H}}\over{\delta q^A }}(x',t) \delta(x,x'),
\label{eq:UnH2}
\end{eqnarray}
which can be recognised as Hamilton's equations in the usual
equal-time sense. This follows from the fact that the Hamiltonian in
equation (\ref{eq:UnA}) 
is by construction independent of any time derivatives and, hence, one can
integrate trivially over $\int dt' \delta(t,t')$.

The weak equality sign is a reminder of the fact that Hamilton's
equations---and hence the actual theory---are not preserved under a
general Poisson bracket. In the
equal-time formalism this presents no problem because the canonical
velocities are only defined externally but, here, they are equally included
in the phase space. As a result, the Poisson bracket between a field
velocitiy and its conjugate momentum can be evaluated to give a
time derivative of the $\delta$-function, which is not the result one
will get if the corresponding Hamilton equation is used to replace the
field velocity before the commutation. Nonetheless, since the theory
is about time evolution only, it is sufficient that Hamilton's
equations are preserved weakly under the Poisson bracket with the
Hamiltonian.

In the unconstrained theory (\ref{eq:UnA}) this follows automatically
from Hamilton's equations and the definition of the general time
Poisson bracket (\ref{eq:UnBr}) without any reference to the specific 
form of the Hamiltonian.  Before checking this explicitly, however, we need to
extend the definition of the Hamiltonian in order to incorporate the
trivial dynamical evolution of the prescribed functions $c^K$, $N$ and $N^i$.  
This is also appropriate for the completeness of the formalism.

\subsection{Incorporating the fixed functions.}
One defines the extended unconstrained action by 
\begin{eqnarray}
&& S[q^A , p_A , {\omega}_K ,\omega , {\omega}_i] = \int d^3dt \bigg(
p_A \dot{q^A} + {\omega_K} \dot{c^K} + {\omega} \dot{N} + {\omega_i}
\dot{N^i} - {\cal H}^{ext} \bigg), 
\nonumber\\
&& {\cal H}^{ext} = N H + N^i H_i + {\omega_K} \dot{c^K} + {\omega} \dot{N}
+ {\omega_i} \dot{N^i}, 
\label{eq:ExtA}
\end{eqnarray}
where the momenta ${\omega}_K$, $\omega$ and $\omega_i$ are defined
through the Poisson bracket relations
\begin{eqnarray}
&& \{ c^K(x,t) , {\omega_L}(x',t') \} = {\delta^K}_L \delta(x,x') \delta(t,t'),
\nonumber\\
&& \{ N(x,t) , {\omega}(x',t') \} = \delta(x,x') \delta(t,t'),
\nonumber\\
&& \{ N^i(x,t) , {\omega_j}(x',t') \} = {\delta^i}_j \delta(x,x')
\delta(t,t').
\label{eq:PbN}
\end{eqnarray}

These momenta are not assumed to have any direct physical significance or
interpretation, and the whole
purpose of their introduction is to allow the time derivative of the
fixed functions to be calculated inside the Poisson bracket
formalism. Restricting our attention to functionals of the
canonical and the fixed variables one gets  
\begin{eqnarray}
&& \{ F(x,t) , \int d^3x' dt' {\cal H}^{ext}(x',t') \} =
{{\delta F} \over {\delta q^A}}(x,t) \{ q^A(x,t), \int
d^3x' dt' {\cal H}^{ext}(x',t') \}    
\nonumber\\
&& + {{\delta F} \over {\delta p_A}}(x,t) \{
p_A(x,t), \int d^3x' dt' {\cal H}^{ext}(x',t') \} + {{\delta F} \over
{\delta c^K}}(x,t) \dot{c^K}(x,t) 
\nonumber\\
&& + {{\delta F} \over
{\delta N}}(x,t) \dot{N}(x,t)  + {{\delta F} \over {\delta N^i}}(x,t) \dot{N^i}(x,t) \simeq \dot{F}(x,t) ,
\label{eq:UnProof}
\end{eqnarray}
which implies that the extended Hamiltonian can be seen as the canonical
representation of the total time derivative operator.

Equivalently one may observe that, when acting on F, the kinematical half
of the extended action  
\begin{equation}
\int d^3dt \bigg(p_A \dot{q^A} + {\omega_K} \dot{c^K} + {\omega}
\dot{N} + {\omega_i} \dot{N^i} \bigg)
\label{eq:kineterms} 
\end{equation}
produces the time derivative of F in the strong sense. A weakly
vanishing result on the other hand arises, by definition, when the
total extended action acts on any F. One concludes that the remaining
half of the action---i.e., the dynamical half corresponding to the
integral of the extended Hamiltonian---produces the total time derivative
of F in the weak sense.

Using either of the above methods, one can prove that
Hamilton's equations are automatically preserved under the dynamical
evolution  of the theory. Indeed, if F is any functional of the
canonical and the fixed variables that vanishes on the constraint surface,
it follows that its total time derivative will also vanish on the same
surface. Since this derivative is weakly equal to the
commutation of F with the integral of the extended Hamiltonian, it
follows that all weakly vanishing functionals remain weakly zero under
this commutation. Choosing these Fs to be Hamilton's
equations themselves shows that the constraint surface is preserved. 
This completes the treatment of systems that are unconstrained in the usual
sense.

\subsection{The constrained Hamiltonian.} 

The extended form of the action, equation (\ref{eq:ExtA}), arises
naturally when the functions $N$ and $N^i$ 
are either constrained canonical variables or acquire the meaning of
Lagrange multipliers. An example of the first case is the history
formulation of general relativity\cite{IKDID}, where one does not use
the Dirac procedure for passing to the Hamiltonian but, instead,
follows the usual Legendre definition without replacing the
non-invertible velocity terms. An example of the second case is the
history formulation of parametrized theories.

We present both these cases in their most general 
form by considering the canonical action
\begin{eqnarray}
&& S[q^A , p_A , N, \omega, N^i, {\omega}_i, {\omega}_K] = \int d^3dt
\bigg( p_A \dot{q^A} + {\omega}_K \dot{c^K} + {\omega} \dot{N} +
{\omega_i} \dot{N^i} - {\cal H} \bigg), 
\nonumber\\
&& {\cal H} = N H + N^i H_i + {\omega}_K \dot{c^K} + {\omega} \dot{N}
+ {\omega_i} \dot{N^i}, 
\label{eq:CA}
\end{eqnarray}
which is now additionally varied with respect to the functions $N$ and
$N^i$. The fields $c^K$ are still treated as fixed.

The variation of (\ref{eq:CA}) leads to the same equations as before, namely 
\begin{eqnarray}
&& \{ S , q^A(x,t) \} \simeq 0 \Leftrightarrow \dot{q^A}(x,t)
\simeq \int d^3x' \bigg(  N {{\delta {H}}\over{\delta p_A }} +
N^i {{\delta {H_i}}\over{\delta p_A }} \bigg) (x',t) \delta(x,x'),  
\label{eq:CcV1}
\\
&& \{ S , p_A(x,t) \} \simeq 0 \Leftrightarrow \dot{p_A}(x,t)
\simeq \int d^3x' \bigg(  N {{\delta {H}}\over{\delta q^A }} +
N^i {{\delta {H_i}}\over{\delta q^A }} \bigg) (x',t) \delta(x,x'),
\label{eq:CcV2}
\\
&& \{ S , c^K(x,t) \} = 0 \Leftrightarrow \dot{c^K}(x,t) =
\dot{c^K}(x,t) \Leftrightarrow 0 = 0 ,
\label{eq:CcV3}
\\
&& \{ S , N(x,t) \} = 0 \Leftrightarrow \dot{N}(x,t) = \dot{N}(x,t)
\Leftrightarrow 0 = 0 ,
\label{eq:CcV4}
\\
&& \{ S , N^i(x,t) \} = 0 \Leftrightarrow \dot{N^i}(x,t) =
\dot{N^i}(x,t) \Leftrightarrow 0 = 0 ,
\label{eq:CcV5}
\end{eqnarray}
subject to the additional equations
\begin{eqnarray}
&& \{ S , {\omega}(x,t) \} \simeq 0 \Leftrightarrow \dot{\omega}(x,t)
\simeq \dot{\omega}(x,t) + {H}(x,t) \Leftrightarrow {H}(x,t) \simeq 0,
\label{eq:CcV6}
\\
&& \{ S , {\omega}_i(x,t) \} \simeq 0 \Leftrightarrow \dot{{\omega}_i}(x,t)
\simeq \dot{{\omega}_i}(x,t) + {H_i}(x,t) 
\Leftrightarrow {H_i}(x,t) \simeq 0,
\label{eq:CcV7}
\end{eqnarray}
arising from the variation of the action with respect to $N$ and $N^i$.

For a functional $F[q^A,p_A,c^K,N,N^i]$ the proof of the previous
section still applies,  
\begin{equation}
\{ F(x,t) , \int d^3x' dt' {\cal H}(x',t') \} \simeq \dot{F}(x,t),
\label{eq:dra}
\end{equation}
with the weak equality refering to Hamilton's equations
(\ref{eq:CcV1}-\ref{eq:CcV2}). It follows that if F is any
functional that vanishes on the surface defined by Hamilton's 
equations, its time derivative will also vanish on this surface,
and by taking F to be Hamilton's equations themselves one can deduce that
(\ref{eq:CcV1}-\ref{eq:CcV5}) are weakly preserved under the dynamical
evolution of the theory. On the other hand, if F vanishes on the
surface defined by the constraint equations
(\ref{eq:CcV6}-\ref{eq:CcV7}), its time derivative will still 
vanish on this surface but, now, it does not follow that this
time derivative will be the one generated by the Hamiltonian of the
theory.

One must ensure that the time derivatives of the fields
calculated by differentiating equations (\ref{eq:CcV6}-\ref{eq:CcV7})
are compatible with the time derivatives of the same fields calculated
from Hamilton's equations. If the constraints 
(\ref{eq:CcV6}-\ref{eq:CcV7}) do not depend on the prescribed fields
$c^K$---which is the case for most of the physical theories---this 
compatibility condition results in the requirement that the algebra of 
$H$ and $H_i$ must close weakly under the general-time Poisson
bracket. Since $H$ and $H_i$ are by construction independent of
any time derivatives, the weak closure of the algebra only refers to
the constraint equations (\ref{eq:CcV6}-\ref{eq:CcV7}).

\section{The evolution postulate.}

\subsection{The inverse procedure and the evolution postulate.}
The aim is to invert the above argument, and derive the general
canonical Hamiltonian of a theory from a set of first principles. The
requirement for these principles to be minimal implies that 
the appropriate starting point of the derivation is the form
(\ref{eq:CA}) of the canonical action. This form, which is the only
prerequisite for the existence of a canonical algebra in the theory,
is valid for both constrained and unconstrained systems, and is
present in both the approaches in \cite{HKT} and \cite{T} as the
so-called ``evolution postulate''. In case that this postulate turns
out to be insufficient to 
determine the theory completely, the plan is that any supplementary
conditions that may be added must be such that the connection between
them remains clear throughout the derivation.

The evolution postulate is to be understood as follows. Initially, one
looks for   
the most general canonical representation of the Hamiltonian that
satisfies the unconstrained version of the postulate, 
\begin{eqnarray}
&& {\partial \over {\partial}t}{q^A}(x,t) \simeq \int d^3x'dt' \{
q^A(x,t) , (N H + N^i H_i)(x',t') \}, 
\label{eq:P1}
\\
&& {\partial \over {\partial}t}{p_A}(x,t) \simeq \int d^3x'dt' \{
p_A(x,t) , (N H + N^i H_i)(x',t') \},
\label{eq:P2}
\end{eqnarray}
where the functions $N$ and $N^i$ can take arbitrary values and the
canonical generators can also depend on some fixed fields $c^K$. It
should be mentioned here that when we say ``can take arbitrary
values'' we essentially mean that the formalism {\it allows} $N$ and $N^i$
to take arbitrary values, although in practice $N$ and $N^i$ will be required
to be positive.

If such a Hamiltonian cannot be found, one resorts to the alternative
possibility of varying the action with respect to the functions $N$
and $N^i$. Equations (\ref{eq:P1}-\ref{eq:P2}) must then be
supplemented by the constraint equations   
\begin{eqnarray}
&& H(x,t) \simeq 0,
\label{eq:P3}
\\
&& {H_i}(x,t) \simeq 0,
\label{eq:P4}
\end{eqnarray}
which have to be preserved under the dynamical evolution of the
theory. This consistency requirement---amounting to the weak closure of the
algebra---is supposed to be included among the evolution postulate
for constrained systems. Note that since the
time derivatives of $N$ and $N^i$ do not appear in the equations of motion
(\ref{eq:P1}-\ref{eq:P4}), $N$ and $N^i$ can still take arbitrary
numerical values. The evolution postulate for both constrained and 
unconstrained systems can then be stated as the
requirement that the canonical action is of the form (\ref{eq:CA})
with $N$ and $N^i$ arbitrary.

\subsection{The evolution postulate in an equivalent form.}

At first sight, conditions (\ref{eq:P1}-\ref{eq:P2}) seem to be
rather too loose for something definite to be drawn out of them. It
seems that the canonical representations can be chosen at will, and
that any constrained theory can be created by just requiring the
closure of the resulting algebra. This view changes radically, however,
when one realizes that the canonical fields have a precise geometric
meaning that has to be respected by the Hamiltonian system. 
In a scalar field theory, for example, the field $\phi(x,t)$ is not
merely a spatial scalar but is also by definition the pull-back of a
spacetime scalar field. Below, the evolution postulate is transformed
to an equivalent condition on the canonical generators that is more
appropriate for the exploitation of this fact.

The functions $N$ and $N^i$ are chosen as the lapse function
and the shift vector, and this will be the case henceforth unless
stated otherwise. Decomposing the time 
derivative operator in equations (\ref{eq:P1}-\ref{eq:P2}) according
to the lapse-shift formula, 
\begin{equation}
{\partial \over {\partial t}} = Nn^{\alpha}
{\partial \over {\partial X^{\alpha}}} + {N^i}{{\cal
X}^{\alpha}}_i {\partial \over {\partial X^{\alpha}}},
\label{eq:dxdt}
\end{equation}
and introducing the momentum ${{\cal P}_{\alpha}}$ conjugate to
the embedding, i.e., 
\begin{equation}
\{ {{\cal X}^{\alpha}}(x,t) , {{\cal P}_{\beta}}(x',t') \} =
{{\delta}^{\alpha}}_{\beta} {\delta}(x,x') {\delta}(t,t')
\label{eq:Pmom}
\end{equation} 
one can bring (\ref{eq:P1}-\ref{eq:P2}) into the form:
\begin{eqnarray}
&& \{ {q^A}(x,t), H(x',t') \} \simeq \{ {q^A}[{\cal
X}(x,t)] , {{\cal P}_{\beta}}(x',t') \} n^{\beta}(x',t')  ,
\label{eq:PP1}
\\
&& \{ {q^A}(x,t), {H_i}(x',t') \} \simeq \{ {q^A}[{\cal
X}(x,t)] , {{\cal P}_{\beta}}(x',t') \} {{{\cal
X}^{\beta}}_i}(x',t') ,  
\label{eq:PP2}
\\ 
&& \{ {p_A}(x,t), H(x',t') \} \simeq \{ {p_A}[{\cal
X}(x,t)] , {{\cal P}_{\beta}}(x',t') \} n^{\beta}(x',t') ,  
\label{eq:PP3}
\\
&& \{ {p_A}(x,t), {H_i}(x',t') \} \simeq \{ {p_A}[{\cal
X}(x,t)] , {{\cal P}_{\beta}}(x',t') \} {{{\cal X}^{\beta}}_i}(x',t'). 
\label{eq:PP4}
\end{eqnarray}
  
Note that the arbitrariness of $N$ and $N^i$ was used to eliminate
the integration.

On the right side of the above equations the explicit dependence of
the fields on the spacetime embedding is fully taken into account. For
the configuration fields this is just the dependence arising from
the definition of the fields as geometric objects in spacetime. For
the conjugate fields the situation is more complicated, and
equation (\ref{eq:P1}) is assumed to be inverted to express the momenta as
functionals of the configuration variables, the lapse, the shift,
and the prescribed fields $c^K$. All the latter have a definite
dependence on the spacetime embedding which is then conveyed to the
conjugate canonical fields.

Equation (\ref{eq:P1}) is always invertible for the momenta because the system
is by construction constrained only in the quantities $N$ and $N^i$ at
the very most. There is one exception to this rule when the action is not
derivable from a spacetime Lagrangean but, instead, is brought
into the form (\ref{eq:CA}) through the introduction of Lagrange
multipliers, as in the case of parametrized theories. Nonetheless, the
momentum can still be defined as a functional of the spacetime
embedding as will be shown elsewhere\cite{IK}.

For the purposes of performing actual calculations, the evolution
postulate is to be used in the following way. Any time derivatives of the
canonical variables that arise on the right side of equations
(\ref{eq:PP1}-\ref{eq:PP4}) are replaced by the original
Hamilton's equations (\ref{eq:P1}-\ref{eq:P2}). When the theory is
unconstrained, this results in a coupled system of four functional
differential equations for $H$ and $H_i$, whose solution---if it
exists---corresponds to the general canonical representation
compatible with the evolution postulate. When the theory is
constrained, on the other hand, the resulting conditions on $H$ and
$H_i$ are not proper differential equations since it is sufficient
that they only hold on the constraint surface (\ref{eq:P3}-\ref{eq:P4}).

If the constraints (\ref{eq:P3}-\ref{eq:P4}) implied that the
canonical variables can not be treated as independent in these
conditions for $H$ and $H_i$, the evolution postulate for constrained
systems would not make any sense at all. However, by construction of
the canonical formalism the constraints must be imposed only {\it
after} the Poisson  brackets have been evaluated. It follows therefore
that---even for constrained systems---the differential equations for
$H$ and $H_i$ must be solved treating the canonical variables as
independent and imposing the constraints (\ref{eq:P3}-\ref{eq:P4})
only at the end. Equivalently, one simply adds on each of the
differential equations an arbitrary term whose value is required to
vanish on the constraint surface. We shall see exactly how this works in the
following section.

Finally note that the replacement of the field
velocities in equations (\ref{eq:PP1}-\ref{eq:PP4}) by the original
and equivalent equations (\ref{eq:P1}-\ref{eq:P2}) does not lead to
cyclic identities as one might have expected. The reason is that---due
to the arbitrariness of $N$ and $N^i$---the latter equations hold in
integrated form while the former hold at every point in space and
time. The information incorporated in these equations is actually so rich 
that it determines the canonical theory completely.

\section{The canonical theory regained.}

That no further assumptions are needed in order to recover a canonical
theory from first principles can be shown in an indirect way, by
starting from the evolution postulate and deriving the additional
postulates of Kucha\v{r} {\it et al}. For constrained systems, it
turns out that these postulates have to be imposed weakly, which is
also predicted from a revised version of Teitelboim's argument. The
new solutions arising from this modification are displayed here in
the case of gravity.

\subsection{Derivation of the reshuffling and ultralocality postulates.}

On the right side of equations (\ref{eq:PP1}-\ref{eq:PP2}) the
configuration fields are treated as functionals of the embedding
relative to which the decomposition of the spacetime theory has been
performed. The reshuffling and ultralocality postulates follow
immediately from equations (\ref{eq:PP1}-\ref{eq:PP2}) once the geometric
meaning of the configuration variable is taken into account. This is
also recognized in \cite{HKT} although, there, the emphasis is given on the
compatibility of the postulates with the dynamical law
(\ref{eq:P1}-\ref{eq:P2}) rather than on the fact that the
postulates are uniquely determined by this law. Referring to the
corresponding comment in section 2, it is only 
because of this fact that the method in \cite{HKT} can be used unambiguously
as an algorithm for finding the physically relevant representations of
a general canonical algebra.

Below, we write down the ultralocality and reshuffling postulates for
the physical examples that one usually considers. The relevant
calculations can be found in the appendix. Note that a strong equality
sign is used, with the understanding that all canonical velocities
have been eliminated through the corresponding Hamilton's equations. 
This is consistent with our general plan, according to which we
originally look for an {\it unconstrained} representation of the
evolution postulate. If a theory is proved to be constrained we will
revise the following equations accordingly.

\paragraph{Scalar field theory.}

The configuration variable is the pullback of a spacetime scalar field,
\begin{equation}
\phi(x,t) = {\phi}[{\cal X}](x,t),
\label{eq:aygo}
\end{equation}
and, as such, is an ultralocal function of the embedding. Equations
(\ref{eq:PP1}-\ref{eq:PP2}) become 
\begin{eqnarray}
&& \{ {\phi}(x,t), H^{\phi}(x',t') \} =
{\phi}_{,{\beta}}(x,t) n^{\beta}(x,t) \delta(x,x') \delta(t,t')
\label{eq:SF1}
\\
&& \{ {\phi}(x,t), {{H^{\phi}}_i}(x',t') \} =
{\phi}_{,i}(x,t) \delta(x,x') \delta(t,t'),
\label{eq:SF2}
\end{eqnarray} 
which can be recognized as the history analogues of the reshuffling
and ultralocality conditions in \cite{K}. Indeed, the 
$\delta(t,t')$ function indicates that the canonical generators are 
independent of the field velocities, the ultralocality of the first
equation implies that the super-Hamiltonian is an ultralocal function
of the momenta, while the form of the second equation ensures that
the super-momentum just reshuffles the data on the hypersurface.

\paragraph{General relativity.}

The configuration variable is the pullback of the spacetime metric,
\begin{equation}
g_{ij}(x,t) = {\gamma}_{{\alpha}\beta}[{\cal X}](x,t) {{\cal
X}^{\alpha}}_i(x,t) {{\cal X}^{\beta}}_j(x,t).
\label{eq:aygo2}
\end{equation}
and equations (\ref{eq:PP1}-\ref{eq:PP2}) result in the 
following conditions on the canonical generators, 
\begin{eqnarray}
&& \{ g_{ij}(x,t),{H^{gr}}_k(x',t') \} =  g_{ki}(x,t)
{\delta}_{,j}(x,x') \delta(t,t') + g_{kj}(x,t) {\delta}_{,i}(x,x')
\delta(t,t')  
\nonumber\\
&& \; \; \; \; \; \; \; \; \; \; \; \; \; \; \; \; \; \; \; \; \; \;
\; \; \; \; \; \; \; \; \; \; \; \; \; +  \; g_{ij,k}(x,t) 
{\delta}(x,x') \delta(t,t'), 
\label{eq:GR1}
\\
&& \{ g_{ij}(x,t),{H^{gr}}(x',t') \} = 2n_{{\alpha};{\beta}}(x,t)
{{\cal X}^{\alpha}}_i(x,t) {{\cal X}^{\beta}}_j(x,t) \delta(x,x') \delta(t,t').
\label{eq:GR2}
\end{eqnarray}
These are indeed equivalent to the reshuffling and ultralocality postulates
(\ref{eq:can1}-\ref{eq:can2}) that are used in \cite{HKT}. 

\paragraph{Deformation and parametrized theories.}

For the theory of hypersurface deformations, the configuration
variable is the embedding itself. Equations
(\ref{eq:PP1}-\ref{eq:PP2}) become 
\begin{eqnarray}
&& \{ {\cal X}^{\alpha}(x,t), H^{D}(x',t') \} =
n^{\alpha}(x,t) \delta(x,x') \delta(t,t'),
\label{eq:HD11}
\\
&& \{ {\cal X}^{\alpha}(x,t), {{H^D}_i}(x',t') \} =
{{\cal X}^{\alpha}}_{i} \delta(x,x') \delta(t,t'),
\label{eq:HD22}
\end{eqnarray} 
which are the reshuffling and ultralocality conditions for the
deformation theory. Putting equations (\ref{eq:HD11}-\ref{eq:HD22}) and
(\ref{eq:SF1}-\ref{eq:SF2}) together, one gets the corresponding
conditions for a parametrized scalar field theory.

\subsection{Derivation of the super-momentum constraint, of the representation
postulate, and of the principle of path independence.}

This is the revised version of Teitelboim's argument that was mentioned in
section 2, so some of the following results can be found in
\cite{T}\cite{HKT}and are 
only stated here for completeness. Besides the revision of the
argument for constrained systems, the other main difference between
this approach and the approach in \cite{T} is that the present
argument does not rely on the principle of path independence but
derives it.

For the representation postulate the derivation starts from the
following two Jacobi identities, 
\begin{eqnarray}
&& \{ \{ H_j(x',t') , F(x'',t'') \} , H_i(x,t) \} + \{ \{ F(x'',t'') ,
H_i(x,t) \} , H_j(x',t') \}
\nonumber\\
&& \; \; \;  \; \; \; \; \; \; \; \; \; \;  \; \; \; \;
\; \; \; \; \; \;  + \{ \{ H_i(x,t)  , H_j(x',t') \} , F(x'',t'') \} = 0,   
\label{eq:CJ1}
\\
&& \{ \{ {H^D}_j(x',t') , F(x'',t'') \} , {H^D}_i(x,t) \} + \{ \{ F(x'',t'') ,
{H^D}_i(x,t) \} , {H^D}_j(x',t') \}
\nonumber\\
&& \;  \; \; \; \; \; \; \; \; \; \;  \; \; \; \;
\; \; \; \; \; \;  + \{ \{ {H^D}_i(x,t)  , {H^D}_j(x',t') \} ,
F(x'',t'') \} = 0, 
\label{eq:DJ1}   
\end{eqnarray}
that hold, respectively, on the canonical and on the deformation history
phase space. The arbitrary functional $F$ depends on both the canonical
variables $q^A$ and $p_A$, while the action of the deformation
generators on these variables is defined as in section 4. The
notation for the normal and tangential projections of $P_{\alpha}$ is
chosen to coincide with the equal-time definitions
(\ref{eq:defo1}-\ref{eq:defo2}).

We consider only the case when the canonical Hamiltonian is
independent of the prescribed fields $c^K$, which is the relevant
case for general relativity. When prescribed fields are present 
in the Hamiltonian the following derivation still applies but depends
on the actual character of the prescribed fields and---for
simplicity---is avoided. An extensive account of such systems can
be found in \cite{Khyp}.

Having restricted $H$, $H_i$ and $F$ to be pure functionals of the
canonical variables, we compare the first terms in the identities
(\ref{eq:CJ1}) and (\ref{eq:DJ1}). The
evolution postulate implies that  
\begin{equation}
\{ H_j(x',t') , F(x'',t'') \} = \{ {H^D}_j(x',t') , F(x'',t'') \} , 
\label{eq:litsa}
\end{equation}
where both brackets depend solely on the canonical variables because
of the restrictions imposed. The use of the strong sign is due to the
replacement of the field velocities, as it has been already explained.
A further application of the evolution postulate then gives 
\begin{equation}
\{ \{ H_j(x',t') , F(x'',t'') \} , H_i(x,t) \} =  \{ \{ {H^D}_j(x',t')
, F(x'',t'') \} , {H^D}_i(x,t) \} , 
\label{eq:litsa2}
\end{equation}
where we have used the fact that Hamilton's equations
are preserved under the commutation with the Hamiltonian.

Repeating this argument when comparing the second terms in the
identities (\ref{eq:CJ1}) and (\ref{eq:DJ1}), we get that
\begin{equation}
\{ \{ F(x'',t'') , H_i(x,t) \} , H_j(x',t') \} = \{ \{ F(x'',t'') ,
{H^D}_i(x,t) \} , {H^D}_j(x',t') \} ,
\label{eq:parak}
\end{equation}
which implies that the remaining terms in the identities should
also be equal, 
\begin{equation}
\{ \{ H_i(x,t)  , H_j(x',t') \} , F(x'',t'') \} = \{ \{ {H^D}_i(x,t)
, {H^D}_j(x',t') \} , F(x'',t'') \} = 0 .
\label{eq:elpis}
\end{equation}

The commutation between the two deformation generators is 
calculated to give the history analogue of the Dirac relation (\ref{eq:D3}), 
\begin{equation}
\{ {H^D}_i(x,t)  , {H^D}_j(x',t') \} =  {H^D}_j(x,t) {\delta}_i(x,x')
\delta(t,t') - (ix \leftrightarrow jx'), 
\label{eq:HDD3}
\end{equation}
and then the evolution postulate is used once more to give an equation 
that holds exclusively on the canonical phase space,
\begin{equation}
\{ \bigg[ \{ H_i(x,t) , H_j(x',t') \} - \bigg( H_j(x,t) {\delta}_{,i}(x,x')
\delta(t,t') - (ix \leftrightarrow jx') \bigg) \bigg] , F(x'',t'') \} = 0.
\label{eq:HCDD3}
\end{equation}
Since it holds for any choice of the functional F, the following 
relation for the super-momenta arises,  
\begin{equation}
\{ {H}_i(x,t)  , {H}_j(x',t') \} =  {H}_j(x,t) {\delta}_i(x,x')
\delta(t,t') + C_{ij}[x,t;x',t'] - (ix \leftrightarrow jx') ,
\label{eq:HD3}
\end{equation}
where $C_{ij}$ is just a constant term.

The same argument can be applied to the mixed Jacobi identities
\begin{eqnarray}
&& \{ \{ H_j(x',t') , F(x'',t'') \} , H(x,t) \} + \{ \{ F(x'',t'') ,
H(x,t) \} , H_j(x',t') \}
\nonumber\\
&& \; \; \;  \; \; \; \; \; \; \; \; \; \;  \; \; \; \;
\; \; \; \; \; \;  + \{ \{ H(x,t)  , H_j(x',t') \} , F(x'',t'') \} = 0,   
\label{eq:CJ2}
\\
&& \{ \{ {H^D}_j(x',t') , F(x'',t'') \} , {H^D}(x,t) \} + \{ \{ F(x'',t'') ,
{H^D}(x,t) \} , {H^D}_j(x',t') \}
\nonumber\\
&& \;  \; \; \; \; \; \; \; \; \; \;  \; \; \; \;
\; \; \; \; \; \;  + \{ \{ {H^D}(x,t)  , {H^D}_j(x',t') \} ,
F(x'',t'') \} = 0, 
\label{eq:DJ2}   
\end{eqnarray}
resulting in the relation
\begin{equation}
\{ {H}(x,t)  , {H}_i(x',t') \} =  {H}(x,t) {\delta}_i(x,x')
\delta(t,t') + {H}_i(x,t) {\delta}(x,x') \delta(t,t') + C_{i}[x,t;x',t'],
\label{eq:HD2}
\end{equation}
with $C_{i}$ being constant.

The situation changes considerably, when the argument is
applied to the remaining identities between the super-Hamiltonians,
\begin{eqnarray}
&& \{ \{ H(x',t') , F(x'',t'') \} , H(x,t) \} + \{ \{ F(x'',t'') ,
H(x,t) \} , H(x',t') \}
\nonumber\\
&& \; \; \;  \; \; \; \; \; \; \; \; \; \;  \; \; \; \;
\; \; \; \; \; \;  + \{ \{ H(x,t)  , H(x',t') \} , F(x'',t'') \} = 0,   
\label{eq:CJ3}
\\
&& \{ \{ {H^D}(x',t') , F(x'',t'') \} , {H^D}(x,t) \} + \{ \{ F(x'',t'') ,
{H^D}(x,t) \} , {H^D}(x',t') \}
\nonumber\\
&& \;  \; \; \; \; \; \; \; \; \; \;  \; \; \; \;
\; \; \; \; \; \;  + \{ \{ {H^D}(x,t)  , {H^D}(x',t') \} ,
F(x'',t'') \} = 0. 
\label{eq:DJ3}   
\end{eqnarray}
This leads to the relation
\begin{equation}
\{ \{ H(x,t)  , H(x',t') \} , F(x'',t'') \} = \{ \{ {H^D}(x,t)  ,
{H^D}(x',t') \} , F(x'',t'') \} ,
\label{eq:HCD1}
\end{equation}
whose left and right side is evaluated on the canonical
and on the deformation phace space, respectively.

Considering the Poisson bracket between the deformation 
generators one has to deal with the fact that the Dirac algebra is not a
genuine Lie algebra but depends explicity on the spatial
metric,
\begin{equation}
\{ {H^D}(x,t)  , {H^D}(x',t') \} = g^{ij}(x,t) {{H^D}_i}(x,t)
\delta_{,j}(x,x') \delta(t,t') - \xxp .
\label{eq:HDD1}
\end{equation}
Since the theory is by assumption independent of any
prescribed fields, it follows that the metric has to be a {\it
canonical} variable in order to appear in equation (\ref{eq:HCD1}).

Using the evolution postulate and the fact that the  
metric is a canonical variable, one writes equation (\ref{eq:HCD1}) exclusively
on the canonical phase space,  
\begin{eqnarray}
&& \{ \bigg[ \{ H(x,t), H(x',t') \} - \bigg( g^{ij}(x,t) {H_i}(x,t)
\delta_{,j}(x,x') \delta(t,t') - \xxp \bigg) \bigg], F(x'',t'') \} 
\nonumber\\
&& = - \bigg( H_i(x,t) {\delta}_{,j}(x,x') \delta(t,t') \{ g^{ij}(x,t) ,
F(x'',t'') \} - \xxp \bigg) . 
\label{eq:111}
\end{eqnarray} 
The term on the right side is the compensation
needed in order for the metric to be taken inside the Poisson brackets
in the canonical phase space.

Because equation (\ref{eq:111}) is a linear first order equation holding for
an arbitrary choice of functional $F$, it cannot be generally satisfied
unless the super-momenta are constrained to vanish,
\begin{equation}
H_i(x,t) \simeq 0.
\label{eq:tralari} 
\end{equation}
The proof follows from the fact that one can expand both sides of
equation (\ref{eq:111}) in terms of the spatial derivatives of the
$\delta$-functions and then always find---due to
the linearity and the actual form of the equation---particular choices
of functionals $F$ that will violate at least one of the terms in the
expansion.

The constraint (\ref{eq:tralari}) leads to
\begin{equation}
\{ \bigg[ \{ H(x,t), H(x',t') \} - \bigg( g^{ij}(x,t) {H_i}(x,t)
\delta_{,j}(x,x') \delta(t,t') - \xxp \bigg) \bigg], F(x'',t'') \} \simeq 0 ,
\label{eq:1110}
\end{equation} 
which must also hold for every choice of functional $F$.
Teitelboim argued\cite{T} that the weak equation
(\ref{eq:1110})---which in the equal-time approach is derived from the
principle of path independence---is enough to imply  
that the expression      
\begin{equation}
\{ H(x,t), H(x',t') \} - \bigg( g^{ij}(x,t) {H_i}(x,t)
\delta_{,j}(x,x') \delta(t,t') - \xxp \bigg) 
\label{eq:lathos}
\end{equation}
vanishes strongly. 
Specifically, he argued that (\ref{eq:lathos}) must not depend on any canonical
variables because, if it did, particular choices of functionals $F$
could always be found to violate equation (\ref{eq:1110}), in a
process similar to the one described above.  The quantity
(\ref{eq:lathos}) should therefore be equal to a constant function,
which is zero\cite{T}  because of the requirement that the
algebra is weakly closed.  
The requirement of closure actually implies that the constant terms
$C_{ij}$ and $C_{i}$ in equations (\ref{eq:HD3}) and (\ref{eq:HD2})
should also be zero\cite{T} and, hence, the history analogue of the strong
Dirac algebra is derived.

This argument is not generally true, however,  because in a
constrained system  one must additionally ensure that all the terms in
equation (\ref{eq:1110}) are well-defined on the constraint
surface. If any of the 
first partial  derivatives of (\ref{eq:lathos}) does not vanish on the
constraint surface Teitelboim's argument can be applied indeed, and
leads to the conclusion that the expression (\ref{eq:lathos}) is
strongly zero. On the other hand, if both partial derivatives of
(\ref{eq:lathos}) vanish weakly, one cannot find  
well-defined choices for functionals $F$ that violate equation
(\ref{eq:1110}) because, to 
do so, would require the first partial derivatives of any such $F$ to
have an infinite value on the constraint surface. Consequently, the
most general expression for the algebra between the super-Hamiltonians
is the weak Dirac relation mentioned in section 2,
\begin{equation}
\{ {H}(x,t)  , {H}(x',t') \} = g^{ij}(x,t) {{H}_i}(x,t)
\delta_{,j}(x,x') \delta(t,t') + G(x,t;x',t') - \xxp  ,
\label{eq:HDDD1}
\end{equation}
where both the first derivatives of $G$ vanish on the constraint
surface (\ref{eq:tralari}). Note that any constant terms are 
absorbed in this definition of $G$.

One now has to go back and re-examine the validity of the
steps that led to equations (\ref{eq:HD3}), (\ref{eq:HD2}) and
(\ref{eq:HDDD1}), taking into account the fact that the 
system is constrained. The only requirement for the consistency of the
previous procedure is the preservation of any weak equality under
commutation with the canonical generators. However, this is already
included in the definition of the evolution postulate for constrained systems 
and, therefore, one simply has to replace any strong equality signs
with weak ones. The complete history analogue of the weak Dirac
algebra (\ref{eq:E1}-\ref{eq:E3}) is therefore obtained, as well as
the weak reshuffling and ultralocality postulates and the rest of the
weak evolution postulate. 
Note that although the term ``weak'' currently refers to the
constraint surface (\ref{eq:tralari}) the arguments that we have used
do not depend on the actual definition of this surface and, therefore,
in case that the super-Hamiltonian is proved to be constrained all our
conclusions will remain valid.

Finally, let us mention again that the path independence of the
dynamical evolution does not need to be assumed separately in the
above argument, but is a consequence of the evolution postulate. 
In particular, one starts from the derived weak Dirac algebra and the evolution
postulate and repeats Teitelboim's argument in the reverse 
order. It follows immediately that the change in the canonical
variables during the dynamical evolution of the theory will be
independent of the path used in their actual evaluation. 
This is of course to be expected when realising that the principle of
path independence is a consequence of the integrability of Hamilton's
equations. The evolution postulate is just another name for these
equations and, hence, any solution of the postulate will
lead automatically 
to a path-independent dynamical evolution.

\subsection{Derivation of the super-Hamiltonian constraint.}

When the representation postulate is imposed in the weak sense, the
super-Hamiltonian constraint does not follow immediately from the
closure of the Dirac algebra---as in \cite{T}---but it is also 
necessary to take into account the actual form of equations
(\ref{eq:PP1}-\ref{eq:PP4}).  We consider these
equations in the case of general relativity or, more accurately, in
the case when the configuration variable is the pullback of the
spacetime metric.

Refering to the corresponding comment at the end of section 4, the most general
form of the weak evolution postulate is the following: 
\begin{eqnarray}
&& \{ {g_{ij}}(x,t), H(x',t') \} =  2 n_{{\alpha};{\beta}}(x,t) {{\cal
X}^{\alpha}}_i(x,t) {{\cal X}^{\beta}}_j(x,t) \delta(x,x')
{\delta}(t,t') 
\nonumber\\
&& \; \; \; \; \; \; \; \; \; \; \; \; \; \; \; \; \; \; \; \; \; \;
\; \; \; \; \; \; \; \; \; \; \; \; + V_{ij}(x,t;x',t'),    
\label{eq:PPGR1}
\\
&& \{ {g_{ij}}(x,t), {H_k}(x',t') \} =  g_{ki}(x,t) {\delta}_{,j}(x,x')
 {\delta}(t,t') + g_{kj}(x,t) {\delta}_{,i}(x,x') {\delta}(t,t') 
\nonumber\\
&& \; \; \; \; \; \; \; \; \; \; \; \; \; \; \; \; \; \; \; \; \; \;
\; \; \; \; \; \; \; \; \; \; \; \;   +
g_{ij,k}(x,t) {\delta}(x,x') {\delta}(t,t') + V_{ijk}(x,t;x',t'),   
\label{eq:PPGR2}
\\ 
&& \{ {p^{ij}}(x,t), H(x',t') \} = \{ {p^{ij}}[{\cal X}(x,t)] ,
{H^D}(x',t') \} + W^{ij}(x,t;x',t'), 
\label{eq:PPGR3}
\\
&& \{ {p^{ij}}(x,t), {H_k}(x',t') \} = \{ {p^{ij}}[{\cal X}(x,t)] ,
{{H^D}_k}(x',t') \}  + {{W^{ij}}_k}(x,t;x',t') .
\label{eq:PPGR4}
\end{eqnarray}

The tensors $V_{ij}$, $V_{ijk}$, $W^{ij}$ and ${W^{ij}}_k$ depend on
the canonical fields and are required to vanish on the constraint
surface $H_i \simeq 0$. Because of the existence of the additional terms,
the general solution of the coupled set
(\ref{eq:PPGR1}-\ref{eq:PPGR4}) cannot be found explicitly. Nevertheless, the
form of the evolution postulate allows some definite conclusions to be drawn,
a part of which can be used to prove that the Hamiltonian is
constrained. A full discussion can be found in \cite{IK}.

The important observation\cite{HKT} is that the conjugate momentum
$p^{ij}$ must be a tensor density of weight one, in order that the form
$p^{ij}{\delta}g_{ij}$ that appears in the canonical action be
coordinate independent. As a result, the Poisson brackets between the 
tangential deformation generator and $p^{ij}$ depend only on the
weight of the latter, and equation (\ref{eq:PPGR4}) becomes
\begin{eqnarray}
&& \{ {p^{ij}}(x,t), {H_k}(x',t') \} = {{\delta}^j}_k {p^{im}}(x,t) 
{\delta}_{,m}(x,x') {\delta}(t,t') + {{\delta}^i}_k {p^{jm}}(x,t) 
{\delta}_{,m}(x,x') {\delta}(t,t') 
\nonumber\\
&& \; \; \; \; \; \; \; \; \; \; \; \; \; \; \; \; \; \; \; \; \; \;
\; \; \; \; \; \; \; \; \; \; \; \;  - {p^{ij}}(x,t)
{\delta}_{,k}(x,x') {\delta}(t,t') - {p^{ij}}_k(x,t) {\delta} (x,x')
{\delta}(t,t') 
\nonumber\\
&& \; \; \; \; \; \; \; \; \; \; \; \; \; \; \; \; \; \; \; \; \; \;
\; \; \; \; \; \; \; \; \; \; \; \;  + {{W^{ij}}_k}(x,t;x',t') . 
\label{eq:GRmom}
\end{eqnarray}

Consider therefore a solution $(H, H_i)$ of the system
(\ref{eq:PPGR1}-\ref{eq:PPGR4}), taking into account equation
(\ref{eq:GRmom}). By the 
assumption of existence of such a solution, the left sides of
equations (\ref{eq:PPGR2}) 
and (\ref{eq:GRmom}) must satisfy the integrability condition
\begin{equation}
\{ \{ g_{ij}(x,t) , {H_k}(x',t') \} , {p^{mn}}(x'',t'') \} = \{ \{
{p^{mn}}(x'',t''), {H_k}(x',t') \} , g_{ij}(x,t)  \} . 
\label{eq:gnwsto}
\end{equation}    
Because the non-vanishing terms in equations (\ref{eq:PPGR2})
and (\ref{eq:GRmom}) are integrable\cite{HKT}, the weakly vanishing terms
in the same equations should also be integrable,
\begin{equation}
\{ V_{ijk}(x,t;x',t') , {p^{mn}}(x'',t'') \} = \{
{{W^{mn}}_k}(x'',t'';x',t'), \{ g_{ij}(x,t)  \}   ,
\label{eq:neo}
\end{equation} 
and, hence, one can always
find some functionals ${H^*}_i$ and $K_i$ satisfying   
\begin{eqnarray}
&& \{ {g_{ij}}(x,t), {{H^*}_k}(x',t') \} = g_{ki}(x,t) {\delta}_{,j}(x,x')
 {\delta}(t,t') + g_{kj}(x,t) {\delta}_{,i}(x,x') {\delta}(t,t') 
\nonumber\\
&& \; \; \; \; \; \; \; \; \; \; \; \; \; \; \; \; \; \; \; \; \; \;
\; \; \; \; \; \; \; \; \; \; \; \;   +
g_{ij,k}(x,t) {\delta}(x,x') {\delta}(t,t')  ,
\label{eq:eid1}
\\
&& \{ {p^{ij}}(x,t), {{H^*}_k}(x',t') \} = {{\delta}^j}_k {p^{im}}(x,t) 
{\delta}_{,m}(x,x') {\delta}(t,t') + {{\delta}^i}_k {p^{jm}}(x,t) 
{\delta}_{,m}(x,x') {\delta}(t,t') 
\nonumber\\
&& \; \; \; \; \; \; \; \; \; \; \; \; \; \; \; \; \; \; \; \; \; \;
\; \; \; \; \; \; \; \; \; \; \; \;  - {p^{ij}}(x,t)
{\delta}_{,k}(x,x') {\delta}(t,t') - {p^{ij}}_k(x,t) {\delta} (x,x')
{\delta}(t,t') ,
\label{eq:eid2}
\\
&& \{ {g_{ij}}(x,t), {{K}_k}(x',t') \} = V_{ijk}(x,t;x',t') ,
\label{eq:eid3}
\\
&& \{ {p^{ij}}(x,t), {{K}_k}(x',t') \} = {{W^{ij}}_k}(x,t;x',t') .
\label{eq:eid4}
\end{eqnarray}

It follows from equations (\ref{eq:eid1}-\ref{eq:eid4}) 
that every solution $H_i$ of the weak evolution postulate can be
written as the sum of two terms,  
\begin{equation}
H_i  = {H^*}_i + K_i .
\label{eq:sum}
\end{equation} 
Furthermore, the form of ${H^*}_i$ is uniquely fixed by equations
(\ref{eq:eid1}) and (\ref{eq:eid2}), and corresponds to the
super-momentum of general relativity\cite{HKT}, 
\begin{equation}
{H^*}_i = {H^{gr}}_i ,
\label{eq:grisstar}
\end{equation} 
written explicitly in equation (\ref{eq:Mm}).

We can now show that the super-Hamiltonian of the theory is
constrained. As in \cite{T}, this follows from the preservation of the
super-momentum constraint under the dynamical evolution, resulting in the
condition    
\begin{equation}
\{ H(x,t) , H_i(x',t') \} \simeq 0.
\label{eq:fanfara}
\end{equation}    
Using equations (\ref{eq:sum}) and (\ref{eq:grisstar}), this
condition can be written as 
\begin{equation}
\{ H(x,t) , \bigg[ {H^{gr}}_i(x',t') + K_i(x',t') \bigg]\} \simeq  0 
\label{eq:weaklymal}
\end{equation}
or, equivalently, as 
\begin{equation} 
\{ H(x,t) , {H^{gr}}_i(x',t') \} \simeq 0 .
\label{eq:weaklymal2}
\end{equation}
To obtain equation (\ref{eq:weaklymal2}) we used equations
(\ref{eq:eid3}-\ref{eq:eid4}) and the fact that 
${{W^{ij}}_k}$ and $V_{ijk}$ vanish on the constraint surface
(\ref{eq:tralari}).

The form (\ref{eq:Mm}) of the gravitational super-momentum is such that the
left side of equation (\ref{eq:weaklymal2}) depends only on the weight of the
super-Hamiltonian---necessarily being one\cite{HKT},
\begin{equation}
\{ H(x,t) , {H^{gr}}_i(x',t') \} =  H(x,t) {\delta}_{,i}(x,x')
{\delta}(t,t') + H_{,i}(x,t) {\delta}(x,x') {\delta}(t,t') ,
\label{eq:}
\end{equation} 
and hence the constraint $H \simeq 0$ is proved. 
Recall that the actual definition of the constraint surface
does not affect the validity of any of the above arguments, and hence
the procedure just described remains consistent under the
additional constraint.

\section{Final comments.}

We have shown that the procedure devised by the authors of \cite{HKT}
corresponds to the requirement that the canonical action is of the form
(\ref{eq:CA}). For unconstrained systems the correspondence is exact,
and the strong reshuffling, ultralocality and representation
postulates determine the form of the canonical theory completely. For
systems subject to constraints the correspondence is not exact unless the
strong postulates are replaced by weak ones, in which case new canonical
representations arise.

Although the understanding of the relationship
between the strong and the weak equations is no longer an
issue---recall that in the revised version no strong equations are
used---a corresponding issue still exists, and concerns the relation
between the ``strong'' and ``weak'' solutions of the evolution postulate. 
In particular, there is need to understand exclusively in terms of the
weak evolution postulate how the standard representation of general
relativity arises, and also to find out if the new representations
are physically equivalent to the standard one. By ``physically
equivalent'' we mean to generate weakly the same equations of motion
and to lead to the same constraint surface.

A preliminary examination of this issue was actually carried out in the
previous section, when proving that the super-Hamiltonian of the theory
is constrained. Indeed, equation (\ref{eq:sum}) shows that the standard
representation of the super-momentum can be derived from the evolution
postulate as the special case $K_i = 0$. In addition, equations
(\ref{eq:PPGR2}) and (\ref{eq:GRmom}) show that 
all solutions $H_i$ generate weakly the same equations of motion,
while---starting from equation (\ref{eq:sum})---it can also be 
shown\cite{IK} that the 
constraints $H_i$ and ${H^{gr}}_i$ imply each other. The
representations $H_i$ and ${H^{gr}}_i$ are therefore physically
equivalent, and the privileged position occupied by ${H^{gr}}_i$ is
merely because the standard description of the system is minimal.

On the other hand, whether the same is true for the representations of
the super-Hamiltonian cannot be said without further
examination. The complication arises 
because of the inversion of equation (\ref{eq:P1}) in order to define
the momenta as functionals of the embedding, and also because of the
replacement of the field velocities on the right side of equations
(\ref{eq:PPGR1}) and (\ref{eq:PPGR3}) by use of
(\ref{eq:P1}-\ref{eq:P2}).  The former  
procedure involves numerous calculations because the inversion can only be 
achieved implicitly, while the latter implies that the right side of
equations (\ref{eq:PPGR1}) and (\ref{eq:PPGR3}) will not be the same for
all representations and, therefore, makes the issue of the physical
equivalence rather unclear. It would be certainly interesting if   
representations could be found that are not equivalent to
the standard super-Hamiltonian, but this possibility is rather remote
considering the restrictions imposed on the spacetime character of any such
representations by Lovelock's theorem\cite{Love}. We hope to be able to say
more about the new representations in the future\cite{IK}.

\section{Acknowledgements.}
My thanks go to C. Isham, K. Garth, C. Anastopoulos and K. Savvidou
for their help.  
I would also like to thank the ``Alexander S. Onassis Public Benefit
Foundation'' for their financial support. 

\begin{appendix}

\section{Poisson brackets in the history phase space.} 
In the following we denote $\delta(x,x') \delta(t,t')$ by $\delta
\delta$,  ${\partial \over {\partial x^i}} \delta(x,x') \delta(t,t')$
by ${\delta}_{,i}$ $\delta$ and ${\partial \over {\partial t}}
\delta(x,x') \delta(t,t')$
 
by $\delta$ $\dot{\delta}$.
If some expressions are calculated at $(x',t')$ they will be simply
primed.

\begin{eqnarray}
&& \{ {\cal X}^{\alpha} , {\cal P}_{\beta}  \} =
{{\delta}^{\alpha}}_{\beta}   \delta \delta
\nonumber\\
&& \{ {\gamma}_{\alpha{\epsilon}}, {\cal P}_{\beta}  \} =
{\gamma}_{\alpha{\epsilon},\beta} \delta \delta
\nonumber\\
&& \{ {\gamma}^{\alpha{\epsilon}}, {\cal P}_{\beta}  \} =
{{\gamma}^{\alpha{\epsilon}}}_{,\beta} \delta \delta 
\nonumber\\
&& \{ {{\delta}^{\alpha}}_{\epsilon}, {\cal P}_{\beta}  \} = 0
\nonumber\\
&& \{ {{\cal X}^{\alpha}}_i, {\cal P}_{\beta}  \} =
{{\delta}^{\alpha}}_{\beta} {\delta}_{,i} \delta
\nonumber\\
&& \{ {{\cal X}_{\alpha}}_i, {\cal P}_{\beta}  \} =
{\gamma}_{\alpha{\beta}} {\delta}_{,i} \delta +
{\gamma}_{\alpha{\mu},\beta} {{\cal X}^{\mu}}_i \delta \delta
\nonumber\\
&& \{ {\dot{\cal X}}^{\alpha} , {\cal P}_{\beta}  \} =
{{\delta}^{\alpha}}_{\beta} \delta \dot{\delta}
\nonumber\\
&& \{ n^{\alpha}, {\cal P}_{\beta}  \} = -n_{\beta} {\cal
X}^{{\alpha}m} {\delta}_{,m} \delta -{1 \over
2}{{\gamma}_{\mu{\nu},\beta}} n^{\mu} n^{\nu} n^{\alpha} \delta \delta
-{{\gamma}_{\mu{\nu},\beta}} n^{\mu} {\gamma}^{{\alpha}\nu} \delta\delta 
\nonumber\\ 
&& \{ n_{\alpha}, {\cal P}_{\beta}  \} = -n_{\beta} {{\cal X}_{\alpha}}^m {\delta}_{,m} \delta -{1 \over2}{{\gamma}_{\mu{\nu},\beta}} n^{\mu} n^{\nu} n_{\alpha} \delta \delta
\nonumber\\
&& \{ g_{ij}, {\cal P}_{\beta}  \} = {{\cal X}_{\beta}}_i
{\delta}_{,j} \delta + {{\cal X}_{\beta}}_j {\delta}_{,i} \delta +
{{\gamma}_{\mu{\nu},\beta}} {{\cal X}^{\mu}}_i {{\cal X}^{\nu}}_j
\delta \delta
\nonumber\\
&& \{ g^{ij}, {\cal P}_{\beta}  \} = -{{\cal X}_{\beta}}^i
g^{jm} {\delta}_{,m} \delta -{{\cal X}_{\beta}}^j
g^{im} {\delta}_{,m} \delta - {{\gamma}_{\mu{\nu},\beta}} {{\cal X}^{\mu}}^i {{\cal X}^{\nu}}^j
\delta \delta
\nonumber\\
&& \{ {{\delta}^i}_j, {\cal P}_{\beta}  \} = 0
\nonumber\\
&& \{ {{\cal X}^{\alpha}}^i, {\cal P}_{\beta}  \} =
-n^{\alpha} n_{\beta} g^{im} {\delta}_{,m} \delta -{{\cal
X}^{\alpha}}^{m} {{\cal X}_{\beta}}^{i} {\delta}_{,m} \delta -
{{\gamma}_{\mu{\nu},\beta}} {{\cal X}^{\alpha}}_m {{\cal X}^{\nu}}^m
{{\cal X}^{\mu}}^i \delta \delta
\nonumber\\
&& \{ {{\cal X}_{\alpha}}^i, {\cal P}_{\beta}  \} = -n_{\alpha} n_{\beta} g^{im} {\delta}_{,m} \delta -{{\cal
X}_{\alpha}}^{m} {{\cal X}_{\beta}}^{i} {\delta}_{,m} \delta -
{{\gamma}_{\mu{\nu},\beta}} n_{\alpha} n^{\nu}
{{\cal X}^{\mu}}^i \delta \delta
\nonumber\\
&& \{ g, {\cal P}_{\beta}  \} = 2g {{\cal X}_{\beta}}^m
{\delta}_{,m} \delta + g {{\gamma}_{\mu{\nu},\beta}} {{\cal X}^{\mu}}_m
{{\cal X}^{\nu}}^m \delta \delta
\nonumber\\
&& \{ N, {\cal P}_{\beta}  \} = -n_{\beta} \delta
\dot{\delta} +n_{\beta} N^m {\delta}_{,m} \delta - {1 \over 2} N 
{{\gamma}_{\mu{\nu},\beta}} n^{\mu} n^{\nu}  \delta \delta 
\nonumber\\
&& \{ N^i, {\cal P}_{\beta}  \} = {{\cal X}_{\beta}}^i
\delta \dot{\delta} + N n_{\beta} g^{im} {\delta}_{,m} \delta - N^m
{{\cal X}_{\beta}}^i {\delta}_{,m} \delta + N
{{\gamma}_{\mu{\nu},\beta}} n^{\mu} {{\cal X}^{\nu}}^i \delta \delta 
\nonumber\\
\label{eq:PB1}
\end{eqnarray}

\end{appendix}

\end{document}